\documentclass{ws-procs9x6}
\usepackage{epsfig}
\usepackage{cite}

\newcommand{\psl}{\mbox{$p$\hspace{-0.4em}\raisebox{0.1ex}{$/$}}}
\newcommand{\vsl}{\mbox{$v$\hspace{-0.5em}\raisebox{0.1ex}{$/$}}}
\def\NP{\hbox{\tiny NP}}
\def\PT{\hbox{\tiny PT}}
\def\PV{\hbox{\tiny PV}}
\def\MSbar{\hbox{\tiny ${\overline{\rm MS}}$}}
\def\tot{\hbox{\tiny tot}}

\begin{document}


\title{inclusive B-decay spectra and IR renormalons}

\author{EINAN GARDI}

\address{Cavendish Laboratory, University of Cambridge\\
Madingley Road, Cambridge, CB3 0HE, UK}

\maketitle

\abstracts{I illustrate the role of infrared renormalons in computing inclusive B-decay spectra. 
I explain the relation between the leading ambiguity in the definition of Sudakov form factor $\sim \exp(N\Lambda/M)$ and that of the pole mass, and show how these ambiguities cancel out between the perturbative and non-perturbative components of the b-quark distribution in 
the meson.}

\section{Introduction}

B-decay physics is gradually turning into a field of precision phenomenology.
Inclusive decay measurements provide some of the most robust tests of the standard model. Classical examples are the rate of $\bar{B}\longrightarrow X_s \gamma$ decays\cite{XsG} and constraints on the unitarity triangle through the measurement of $V_{ub}$ from charmless semileptonic decays\cite{SL}.
 
The advantage of inclusive measurements over exclusive ones is that the corresponding 
theoretical predictions are, to large extent, free of hadronic uncertainties. 
QCD corrections to total decay rates are dominated by short distance scales, of order of the heavy-quark mass $m$, and are therefore primarily perturbative. Confinement effects appear as power corrections in $\Lambda/m$. Moreover, the Operator Product Expansion (OPE) allows one to estimate these power corrections by relating them to specific matrix elements of local operators between B-meson states, which are defined in the infinite--mass limit in the framework of the heavy-quark effective theory (HQET)\cite{Neubert:1996wg}. 
These matrix elements can either be computed on the lattice or extracted from experimental data.

In reality, however, experiments cannot perform completely inclusive measurements.
Precise measurements are restricted to certain kinematic regions where the background is sufficiently low. The experimentally accessible region in $\bar{B}\longrightarrow X_s \gamma$ is where the photon energy $E_{\gamma}$ in the B rest frame is close to its maximal possible value, $M/2$, ($M$ is the B meson mass), or, equivalently, $x\equiv 2E_{\gamma}/M$ is near $1$, which is the endpoint. 
Similarly, the accessible region in the CKM-suppressed $B\longrightarrow X_u l \bar{\nu}$ decay is where the lepton energy fraction is near maximal, or where the invariant mass of the hadronic system is small.
Out of this region this decay mode is completely overshadowed by the decay into charm.

As a consequence, precision phenomenology must rely on detailed theoretical understanding of the 
spectrum\cite{Luke:2003nu}. Of particular importance is the spectrum near the endpoint. It turn out, however, that the endpoint region is theoretically much harder to access as both the perturbative expansion and the OPE break down there. 
In the large-$x$ region gluon emission is restricted to be soft or collinear to the light-quark jet. While the associated singularities cancel with virtual corrections (decay spectra being infrared and collinear safe) large Sudakov logarithms of $(1-x)$ appear in the expansion, which must therefore be resummed.
Moreover, the OPE breaks down since the hierarchy between operators scaling with different powers of the mass is lost when $(1-x)M$ become as small as the QCD scale. Physically this reflects the fact that the spectrum in the endpoint region is driven by the dynamics of the light degrees-of-freedom in the meson. 

The lightcone-momentum distribution\footnote{We shall define this function in full QCD, and call it Quark Distribution Function (QDF). This should be distinguished from the common practice to define it directly in the HQET, where the name ``Shape Function'' is often used.} of the b-quark in the B-meson has a particularly important role in the endpoint region~\cite{Bigi:1993fe,Manohar:1993qn,Neubert:1993ch,Falk:1993vb,Neubert:1993um,Bigi:1993ex,Mannel:1994pm,Kagan:1998ym,Korchemsky:1994jb}. It has been shown that up to subleading corrections ${\mathcal O}(\Lambda/m)$ the physical spectrum can be obtained as the convolution between a perturbatively--calculable coefficient function and the QDF, where the latter essentially determines the 
shape for $x\longrightarrow 1$.
The key point is that the QDF is a property of the B meson, not of the particular decay mode considered, so it can be measured in one decay and used in another.  
Moreover, a systematic analysis of the QDF in the HQET highlights the significance of a few specific parameters which constitute the first few moments of this function: most importantly $\bar{\Lambda}\equiv M-m$, the difference between the meson mass and the quark pole mass, and then $\lambda_1$ corresponding to the kinetic energy of the b quark in the meson. 

Nevertheless, the dependence of theoretical predictions for the spectra on the QDF is still a major source of uncertainty. 
Apart from identifying its first few moments, very little is known about this function, so the phenomenology of decay spectra in the immediate vicinity of the endpoint ($x\longrightarrow 1$) remains, to large extent, model dependent. 
On the other hand, successful precision phenomenology can well be expected for more moderate (yet large) $x$ values, corresponding to the region where the distribution peaks. 
Here the main obstacle has been in combining\cite{Bigi:1993ex,Bosch:2004th} perturbative Sudakov effects with the HQET-based non-perturbative treatment discussed above.

It has recently been shown\cite{BDK} that the resolution of this problem is firmly connected
with infrared renormalons (for general review of renormalons see\cite{renormalons}). 
Since the formulation of the HQET as well as the perturbative calculation of decay spectra rely on the concept of an on-shell heavy quark, both ingredients suffer from renormalon ambiguities. These ambiguities cancel out, of course, in the physical spectra. It is therefore useful to trace\footnote{This can be understood in analogy with factorization scale dependence, the main difference being that here the interest is in power terms.} the precise cancellation of ambiguities: 
the use of the HQET brings about dependence on the quark pole mass, which has a linear renormalon ambiguity\cite{Bigi:1994em,Beneke:1994sw,BBZ,Neubert:1994wq}. This ambiguity cancels against the leading renormalon ambiguity in the Sudakov exponent\cite{BDK}.  
In order to achieve power-like separation between perturbative and non-perturbative contributions to decay spectra, one must therefore compute the Sudakov exponent as an asymptotic expansion, thus replacing the standard Sudakov resummation with fixed logarithmic accuracy by Dressed Gluon Exponentiation (DGE)\cite{Gardi:2001ny,Gardi:2001di,Gardi:2002bk,Gardi:2002xm,CG,Gardi:2003iv,BDK}. 

In what follows we illustrate the role of renormalons in the QCD description of decay spectra. 
We begin by briefly reviewing the HQET analysis for the QDF where we identify dependence on the quark pole mass. We recall that the pole mass suffers from an infrared renormalon ambiguity and show how this affects the QDF\cite{BDK}. We then consider inclusive B-meson decays within perturbation theory, review the relevant results on large-$x$ factorization and Sudakov resummation\cite{Korchemsky:1994jb}, and then show that 
renormalon ambiguities appear in the Sudakov exponent\cite{BDK}, which, we emphasize, is a general phenomenon rather than a peculiarity of B decays. 
Finally, we combine the perturbative and non-perturbative ingredients recovering an unambiguous answer for the QDF in the meson and consequently for decay spectra. We conclude by shortly discussing the implications for precision phenomenology in inclusive decays.  

\section{Heavy-quark effective theory and the QDF} 

We define the QDF $f(z;\mu)$ as the Fourier transform of the forward hadronic matrix element of two heavy--quarks fields on the lightcone (\hbox{$y^2=0$}):
\begin{eqnarray}
f(z;\mu)=\frac{1}{4\pi}\int_{-\infty}^{\infty} d y^-\,{\rm e}^{iz p_B^+ y^-}
\left\langle B(p_B)\right\vert \bar{\Psi}(0)\gamma^+\Psi(y)\left\vert B(p_B)\right\rangle_{\mu}
\label{f},
\end{eqnarray}
where a path-ordered exponential between the fields is understood, $p_B$ is the B-meson momentum ($p_B^2=M^2$), $z$ is the fraction of the ``+'' momentum component carried by the b-quark field
and $\mu$ is the renormalization scale of the operator. 
Decay spectra can be computed as a convolution between a perturbatively calculable coefficient function and $f(z;\mu)$. Let us first analyze $f(z;\mu)$ non-perturbatively --- we denote it $f_{\NP}(z)$ --- suppressing any perturbative corrections. These will be recovered later on. 

Since the $b$-quark mass is large, the heavy quark is not far from its mass shell.  This observation is the basis of the HQET. The momentum of the heavy quark is $p=mv+k$ where $v$ is the hadron four velocity, $v\equiv p_B/M$, and $k$ is a residual momentum, $|k|\ll m$. 
The effective field is defined by scaling out the dependence on the quark mass: 
$h_v(x)={\rm e}^{im v\cdot x}\,\frac12(1+\vsl)\,\Psi(x)$. 
It then follows from the definition (\ref{f}) that in the heavy-quark 
limit\cite{Bigi:1993fe,Manohar:1993qn,Neubert:1993ch,Falk:1993vb,Neubert:1993um,Bigi:1993ex,Mannel:1994pm,Kagan:1998ym,Korchemsky:1994jb,Bosch:2004th,BDK}
\begin{eqnarray}
\label{shape_function}
\hspace*{-10pt}
&&\int_0^1dz f_{\NP}(z){\rm e}^{-iv\cdot y \,(\bar{\Lambda}-(1-z)M)}
=
\frac{1}{2M}\left\langle B(Mv)\right\vert \bar{h}_v(0)
h_v(y^{-})\left\vert B(Mv)\right\rangle \nonumber \\
&&\hspace*{40pt}
=1+\frac{f_2}{2!}(-iv\cdot y)^2+\frac{f_3}{3!}(-iv\cdot y)^3+\cdots \equiv {\mathcal F}(-iv\cdot y),
\end{eqnarray}
where we inverted the Fourier transform and defined $\bar{\Lambda}\equiv M-m$;  
in the second line we expanded the lightcone operator in the HQET in terms of local operators, where e.g.
\begin{eqnarray}
\hspace*{-15pt}
-3f_2&\equiv&\lambda_1\equiv \,\frac{1}{2M}{\left\langle B(Mv)\left\vert \bar{h}_v(0)
(g_{\mu\nu}-v_{\mu}v_{\nu})
iD^{\mu}iD^{\nu} \,h_v(0)\right\vert
B(Mv)\right\rangle}.
\end{eqnarray}
The HQET matrix elements $f_n\sim {\mathcal O}(\Lambda^n)$ do not depend on the
definition of the mass.
On the other hand the vanishing of the linear term in $(-iv\cdot y)$, $f_1=0$, (and the absence of additional, mass dependent terms in front of higher powers of $(-iv\cdot y)$) 
in the second line of Eq.~(\ref{shape_function}) are due to the HQET equation of motion for the heavy quark. Thus Eq.~(\ref{shape_function}) {\em  relies} on using the pole mass to define the HQET\footnote{The use of the pole mass in the field redefinition can be avoided if a residual mass term is introduced. This, however, does not change any of the conclusions\cite{BDK}.}.

Expanding the exponential on the l.h.s of Eq.~(\ref{shape_function}) we obtain:
\begin{equation}
\label{f_n}
\hspace*{-5pt}
f_n=\int_0^1f_{\NP}(z)\,\left(\bar{\Lambda}-(1-z)M\right)^n,
\end{equation}
so the moments are fixed by the local matrix elements in the HQET. 
Mellin moments are defined by 
\begin{equation}
F_N^{\NP}=\int_0^1dz z^{N-1} f_{\NP}(z).
\end{equation}
For the first few Mellin moments we have $F_1^{\NP}=1$ and 
\begin{eqnarray}
\label{first_few_moments}
F_2^{\NP}=\frac{m}{M}\,;\quad
F_3^{\NP}=\left(\frac{m}{M}\right)^2-\frac13\frac{\lambda_1}{M^2}\,;\quad
F_4^{\NP}=\left(\frac{m}{M}\right)^3-\frac{\lambda_1 m}{M^3}+\frac{f_3}{M^3}.
\end{eqnarray}
It is apparent that all the moments depend on the quark pole mass. They satisfy
$M\,{dF_N^{\NP}}/{dm}=({N-1})\, F_{N-1}^{\NP}$.
At large $N$ they are given by\cite{BDK}
\begin{eqnarray}
\label{F_NP}
F_N^{\NP}={\mathrm e}^{-(N-1)\bar{\Lambda}/M}{\mathcal F}((N-1)/M)\,+\,{\mathcal O}(1/N),
\end{eqnarray}
where the exponential factor depends on the pole mass through $\bar{\Lambda}$ while ${\mathcal F}((N-1)/M)$, defined in (\ref{shape_function}), is entirely quark-mass independent. Note that large $N$ corresponds to asymptotically large lightcone separations.

\section{IR renormalon ambiguity in the pole mass}

The result of the previous section indicates inherent dependence of the non-perturbative component of the QDF in the heavy-quark limit on $\bar{\Lambda}=M-m$, or, equivalently, on the pole mass $m$. We recall that the pole mass is defined in perturbation theory by requiring that
the inverse quark propagator $\psl -m_{\MSbar} -\Sigma(p,m_{\MSbar})$ vanishes at $p^2=m^2$. At any given order in
$\alpha_s$ one can solve the resulting equation obtaining a unique relation between the pole mass and the  $\overline{\rm MS}$ mass (or any other renormalized short-distance mass).
However, when considered to power accuracy this definition remains ambiguous\cite{Bigi:1994em,Beneke:1994sw,BBZ,Neubert:1994wq}. 
The on-shell condition brings about linear sensitivity to long-distance scales. In the perturbative expansion (in schemes such as $\overline{\rm MS}$) this sensitivity translates into non-alternating factorial divergence making the sum of the series ambiguous --- the well known infrared renormalon.

Specifically, with a single dressed gluon --- thus to leading order in the large-$\beta_0$ limit --- the relation between the pole mass and $m_{\MSbar}$ is given by the following Borel sum\cite{Beneke:1994sw}:
\begin{equation}
\frac{m}{m_{\MSbar}}\!=\!1-\frac{C_F}{\beta_0}\int_0^{\infty}\!\!\!\!
du \left(\frac{\Lambda^2}{m_{\MSbar}^2}\right)^{\!u}
 \left[ \frac{3{\rm e}^{\frac{5}{3}u}(1-u)\Gamma(u){ \Gamma(1-2u)}}{2\Gamma(3-u)}-\frac{3}{4u}+R_{\Sigma_1}(u)\right],
\label{mass_rel}
\end{equation}
where $R_{\Sigma_1}(u)$ is free of singularities and $\beta_0=\frac{11}{12}C_A-\frac16 N_f$. 
The singularity of the integrand 
at $u=\frac12$ translates into an ${\mathcal O}(\Lambda/m)$ ambiguity, which directly affects the QDF moments $F_N^{\NP}$. 

As one would expect, the pole-mass renormalon ambiguity cancels out whenever the pole mass is used to compute an observable quantity. A well-known example\cite{BBZ} is the calculation of the total semileptonic decay rate, which explicitly depends on the fifth power of the mass. Another example is the total energy of quarkonia\cite{Hoang:1998nz}.
Here we review this cancellation for the QDF in the meson and consequently for decay spectra\cite{BDK}.

\section{Large-$x$ factorization in B decay\label{Factor}}

Let us now consider B-decay spectra in perturbation theory. By taking the initial state to be an on-shell b quark we neglect non-perturbative effects associated with the meson structure. The decay rate is infrared and collinear safe, so the partonic calculation yields finite perturbative expansion when expressed in terms of the renormalized coupling and mass. It should be noted, however, that this finiteness is owing to cancellation of logarithmic singularities between real and virtual corrections. As usual, the singularity leaves a trace in the form of Sudakov logarithms of $(1-x)$. Since such logarithms appear at any order in the perturbative expansion, they must be resummed. 

\begin{figure}[t]
\begin{center}\vskip -20pt
\epsfig{height=4.1cm,angle=0,width=8.0truecm,file=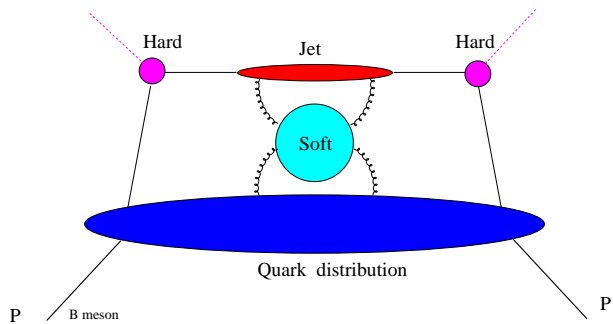}
\caption{Large-$x$ factorization of inclusive decays into soft ($m/N$), jet ($m^2/N$) and hard ($m$) functions.\label{fact}}
\end{center}
\end{figure}

The resummation of Sudakov logarithms takes the form of exponentiation in Mellin space. This is a consequence 
of the factorization property of QCD matrix elements in the soft and collinear limits together with the factorization of phase space~\cite{CSS,CT_CTTW,Korchemsky:1994jb}. Up to ${\mathcal O}(1/N)$ corrections the perturbative expansion of inclusive decay spectra can be written in Mellin space as a product of three functions\cite{Korchemsky:1994jb}: a soft function depending on $m/N$, a jet function depending on $m^2/N$ and a hard function depending on $m$ --- see Fig.~\ref{fact}.

Furthermore, the resummation can be formulated as DGE, resumming running-coupling effects in the Sudakov exponent.
The result is most conveniently expressed as a Borel sum.   
Explicitly, for\footnote{The corresponding formula for the semileptonic decay appears in\cite{BDK}.} $\bar{B}\longrightarrow X_s \gamma$ we have\cite{BDK}: 
\begin{eqnarray}
M_N^{\PT}\equiv \int_0^1\!\!\!
dx\,\frac{1}{\Gamma^{\PT}_{\tot}}\frac{d\Gamma^{\PT}}{dx}x^{N-1}=C_N(m)J_N(m;\mu)S_N(m;\mu)+{\mathcal O}({1}/{N}),
\label{gen_fact_form}
\end{eqnarray}
where 
\begin{eqnarray}
\label{Soft}
S_N(m;\mu)&=&\exp \bigg\{
\frac{C_F}{\beta_0}\int_0^{\infty}\frac{du}{u} \,T(u)\,
\left(\frac{\Lambda^2}{m^2}\right)^u\,\times\\
\nonumber &&\hspace*{30pt} \bigg[
B_{\mathcal S}(u) \Gamma(-2u)\left(N^{2u}-1\right)
 + \left(\frac{m^2}{\mu^2}\right)^u
B_{\mathcal A}(u)\ln N \bigg] \bigg\},\\ \label{Jet}
J_N(m;\mu)&=&\exp \bigg\{
\frac{C_F}{\beta_0}\int_0^{\infty}\frac{du}{u} \,T(u)\,
\left(\frac{\Lambda^2}{m^2}\right)^u\,\times\\
\nonumber &&\hspace*{30pt} \bigg[ -\,B_{\mathcal J}(u) \Gamma(-u)\left(N^{u}-1\right)
-\left(\frac{m^2}{\mu^2}\right)^u
B_{\mathcal A}(u)\ln N
 \bigg] \bigg\},
\end{eqnarray}
Here $S_N(m;\mu)$ and $J_N(m;\mu)$ are the soft and jet functions, respectively\footnote{$T(u)$ depends\cite{BDK} on the approximation used for the $\beta$ function; for one-loop running coupling~\hbox{$T(u)\equiv 1$}.}. These functions were both normalized to unity --- the exponents vanish at $N=1$ --- so they acquire dependence on the hard scale.
$B_{\mathcal S}(u)$, $B_{\mathcal J}(u)$ and $B_{\mathcal A}(u)$ are Borel representations of anomalous dimensions of the soft, jet and cusp functions, respectively. In the large-$\beta_0$ limit
\begin{eqnarray}
B_{\mathcal S}(u)&=&{\rm e}^{cu} \,(1-u)\,+{\mathcal O}(1/\beta_0), \label{BS} \\ 
B_{\mathcal J}(u)&=&\frac12 {\rm e}^{cu}
\left(\frac{1}{1-u}+\frac{2}{2-u}\right) \frac{\sin\pi
u}{\pi u} +{\mathcal O}(1/\beta_0). \label{BJ}
\end{eqnarray}
where $c=5/3$ in ${\overline {\rm MS}}$.
Beyond this limit the anomalous dimensions are known only as an expansion in $u$ (through NNLO).
Terms that are subleading in $1/\beta_0$, which appear first at ${\mathcal O}(u^1)$, 
are not small in QCD.
The advantage of the large-$\beta_0$ limit, where an analytic function is known, is that it allows one to verify the {\em exact} cancellation of renormalon ambiguities.

\begin{figure}[t]
\begin{center}\vskip -20pt
\hspace*{-30pt}\epsfig{height=4.1cm,angle=0,width=8.0truecm,file=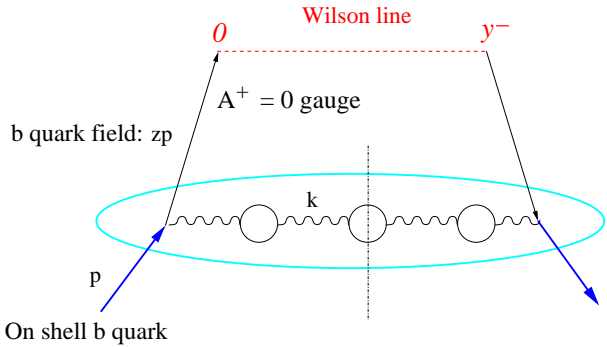}
\caption{Process-independent calculation of the QDF in an on-shell heavy quark in the large-$\beta_0$ limit: the gluon is dressed by any number of fermion-loop insertions and then
$N_f\longrightarrow -6\beta_0$. In the $A^+=0$ axial gauge only this diagram contributes.\label{QDF_ren}}
\end{center}
\end{figure}

The soft function is the large-$N$ limit of the lightcone momentum distribution of a b-quark field in an on-shell b quark. In can be computed in a process-independent manner from the QDF definition (\ref{f}), replacing the external meson states $\left\vert B(P_B)\right\rangle$ by on-shell quark states $\left\vert b(p)\right\rangle$. For example, the large-$\beta_0$ limit result of Eqs. (\ref{Soft}) and (\ref{BS}) can be obtained from the diagram of Fig.~\ref{QDF_ren}. It is related by crossing to the perturbative heavy-quark fragmentation function analyzed in\cite{CG}.

The jet function describes the radiation associated with an unresolved jet of invariant mass $m^2/N$. It is a universal object appearing in many observables including deep inelastic structure functions\cite{Gardi:2001di,Gardi:2002bk,Gardi:2002xm}, 
single-particle inclusive cross sections \cite{Gardi:2001di,CG} and event-shape distribution\cite{Gardi:2001ny,Gardi:2003iv,Berger:2004xf}.

In both the soft and jet functions there are renormalon ambiguities owing to the singularities of $\Gamma(-2u)$ and $\Gamma(-u)$, respectively. They appear as a result of integrating over the longitudinal momentum fraction near the endpoint and reflect the sensitivity of the exponent to large-distance scales through the running of the coupling. The ambiguity indicates the presence of non-perturbative power corrections at the exponent for each of the functions: powers of $N\Lambda/m$ in the soft function and powers of $N\Lambda^2/m^2$ in the jet function. 

\section{Cancellation of renormalon ambiguities in the exponent\label{Cancl}}

When considered to power accuracy, the perturbative soft function of Eq.~(\ref{Soft}) becomes ambiguous. This is not surprising since its definition involves the on-shell quark state $\left\vert b(p)\right\rangle$. Its non-perturbative analog, the QDF in the meson defined in Eq.~(\ref{f}), should be well defined. Yet, at large~$N$ these two functions differ just by (an infinite set of) non-perturbative power corrections on the scale~$M/N$:
\begin{equation}
\label{F_N}
F_N(M;\mu) =S_N(m;\mu)F_N^{\NP},
\end{equation}
where\footnote{We systematically neglect ${\mathcal O} (1/N)$, or, equivalently, ${\mathcal O} (\Lambda/M)$ effects.
Eq.~(\ref{F_N}) does not hold for small $N$.} $F_N^{\NP}$ is given by (\ref{F_NP}). 
The corresponding non-perturbative large-$x$ factorization in B-meson decay is:
\begin{eqnarray}
M_N\equiv \int_0^1\!\!\!
dx\,\frac{1}{\Gamma_{\tot}}\frac{d\Gamma}{dx}x^{N-1}=C_N(m)J_N(m;\mu)F_N(M;\mu)+{\mathcal O}({1}/{N}),
\label{gen_fact_form_NP}
\end{eqnarray} 
where the sole difference from the perturbative formula (\ref{gen_fact_form}) is the replacement of the QDF in the quark, $S_N(m;\mu)$, by that in the meson, $F_N(M;\mu)$.

Since the QDF $F_N(M;\mu)$ directly enters the measurable moments $M_N$, it {\em must be} well defined. 
This will be the case only if renormalon ambiguities cancel in (\ref{F_N}) between the perturbative and non-perturbative components. Indeed, such cancellation is expected because both $S_N(m;\mu)$ and $F_N^{\NP}$ involve the concept of an on-shell heavy quark, while $F_N(M;\mu)$ does not. Putting together (\ref{F_NP}) and (\ref{Soft}) we obtain\cite{BDK}:
\begin{eqnarray}
\label{F_NP_cancl}
F_N(M;\mu)&=&{\mathcal F}((N-1)/M)\,\exp \bigg\{-\frac{(N-1)\bar{\Lambda}}{M}+
\frac{C_F}{\beta_0}\int_0^{\infty}\frac{du}{u} T(u)
\left(\frac{\Lambda^2}{m^2}\right)^u\nonumber \\
 &&\hspace*{-40pt} \times \bigg[
B_{\mathcal S}(u) \Gamma(-2u)\left(N^{2u}-1\right)
 + \left(\frac{m^2}{\mu^2}\right)^u
B_{\mathcal A}(u)\ln N \bigg] \bigg\} +\,{\mathcal O}(1/N),
\end{eqnarray}
which can be explicitly verified to be free of $u=\frac12$ ambiguities in the large-$\beta_0$ limit by substituting $\bar{\Lambda}=M-m$ and using Eqs.~(\ref{mass_rel}) and (\ref{BS}). 
  
\section{Prospects for precision phenomenology}

The results of Secs.~\ref{Factor} and \ref{Cancl} have direct implications for the calculation of inclusive decay spectra: they open up the way for consistent {\em power-like} separation between perturbative and non-perturbative contributions depending on $N\Lambda/M$.
QCD predictions for decay spectra in the peak region require Sudakov resummation
as well as non-perturbative corrections depending on the meson structure. However, conventional 
Sudakov resummation with fixed logarithmic accuracy 
does not deal with the problem of separation between perturbative and non-perturbative contributions (the immediate price is Landau singularities). The perturbative coefficients get significant contributions from small momentum scales, contributions that increase with increasing logarithmic accuracy\cite{Gardi:2001di}. 
DGE addressed the source of this problem. Using Borel summation --- with a principal-value (PV) prescription for example --- it systematically separates non-perturbative power-like terms in the Sudakov exponent from perturbative contributions. This procedure uniquely defines the non-perturbative 
power terms ---
this is precisely the meaning of Eq.~(\ref{F_NP_cancl}): consider for simplicity the hypothetical case where ${\mathcal F}\simeq 1$ so $f_{\NP}(z)\simeq \delta(z-{m}/{M})$. The shape of the QDF is then determined by the perturbative Sudakov form factor; it is just {\em shifted} non-perturbatively by $\bar{\Lambda}/M$ toward smaller $z$ values 
(in the general case ${\mathcal F}$ leads to some smearing). 
Precise control of this shift is, of course, crucial; an ambiguity of order $\Lambda$ in $\bar{\Lambda}$ would be 
a catastrophe. However, based on Eq.~(\ref{F_NP_cancl}) $\bar{\Lambda}$ is uniquely fixed: if the principal value of the Borel sum is used to define the perturbative Sudakov exponent, the {\em same prescription} must be used to relate the pole mass to any (well measured) short-distance mass when computing $\bar{\Lambda}$, so $\bar{\Lambda}=M-m_{\PV}$.     

It should be emphasized that quantitative control of power-like contributions by means of 
Borel summation requires more information than available either 
from fixed-order calculations of the anomalous dimensions or from the 
large-$\beta_0$ limit alone. 
For example, the value of $B_{\mathcal S}(u)$ near $u=\frac12$ becomes relevant. 
While challenging, this question can still be addressed within perturbation theory.

\section*{Acknowledgments} 
I would like to thank Gregory Korchemksy for illuminating discussions.  This work
is supported by a European Community Marie Curie Fellowship, HPMF-CT-2002-02112.



\begin{thebibliography}{0}



\bibitem{XsG}
R.~Barate {\it et al.}  [ALEPH Collaboration],
{\em Phys.\ Lett.\ } {\bf B429} (1998) 169;
%
S.~Chen {\it et al.}  [CLEO Collaboration],
{\em Phys.\ Rev.\ Lett.\  } {\bf 87} (2001) 251807
[hep-ex/0108032];
%
B.~Aubert {\it et al.}  [BaBar Collaboration],
[hep-ex/0207076];
%
P.~Koppenburg {\it et al.}  [Belle Collaboration],
[hep-ex/0403004].


\bibitem{SL}
A.~Bornheim {\it et al.}  [CLEO Collaboration],
{\em Phys.\ Rev.\ Lett.\ } {\bf 88} (2002) 231803
[hep-ex/0202019];
%
B.~Aubert {\it et al.}  [BaBar Collaboration],
{\em Phys.\ Rev.\ Lett.\ } {\bf 92} (2004) 071802
[hep-ex/0307062];
%
B.~Aubert {\it et al.}  [BaBar Collaboration],
{\em Phys.\ Rev.\ } {\bf D69}, 111104 (2004)
[hep-ex/0403030];
%
H.~Kakuno {\it et al.}  [Belle Collaboration],
{\em Phys.\ Rev.\ Lett.\ }  {\bf 92} (2004) 101801
[hep-ex/0311048].



\bibitem{Neubert:1996wg}
M.~Neubert,
``Heavy-quark effective theory,''
hep-ph/9610266.



\bibitem{Luke:2003nu}
M.~Luke,
``Applications of the heavy quark expansion: $|$V(ub)$|$ and spectral
 moments,''
eConf {\bf C0304052} (2003) WG107
[hep-ph/0307378].




\bibitem{Bigi:1993fe}
I.~I.~Y.~Bigi, M.~A.~Shifman, N.~G.~Uraltsev and A.~I.~Vainshtein,
{\em Phys. Rev. Lett.}  {\bf 71} (1993) 496
[hep-ph/9304225].

\bibitem{Manohar:1993qn}
A.~V.~Manohar and M.~B.~Wise,
{\em Phys. Rev.} D {\bf 49} (1994) 1310
[hep-ph/9308246].

\bibitem{Neubert:1993ch}
M.~Neubert,
{\em Phys. Rev.} {\bf D49} (1994) 3392
[hep-ph/9311325].

\bibitem{Falk:1993vb}
A.~F.~Falk, E.~Jenkins, A.~V.~Manohar and M.~B.~Wise,
{\em Phys. Rev.} D {\bf 49} (1994) 4553
[hep-ph/9312306].

\bibitem{Neubert:1993um}
M.~Neubert,
{\em Phys.\ Rev.} {\bf D49} (1994) 4623
[hep-ph/9312311].

\bibitem{Bigi:1993ex}
I.~I.~Bigi, M.~A.~Shifman, N.~G.~Uraltsev and A.~I.~Vainshtein,
{\em Int. J. Mod. Phys.} {\bf A9} (1994) 2467
[hep-ph/9312359].


\bibitem{Mannel:1994pm}
T.~Mannel and M.~Neubert,
{\em Phys. Rev.} {\bf D50} (1994) 2037
[hep-ph/9402288].

\bibitem{Kagan:1998ym}
A.~L.~Kagan and M.~Neubert,
{\em Eur. Phys. J.} {\bf C7} (1999) 5
[hep-ph/9805303].




\bibitem{Korchemsky:1994jb}   
G.~P.~Korchemsky and G.~Sterman,
{\em Phys. Lett.}  {\bf B340} (1994) 96
[hep-ph/9407344].


\bibitem{Bosch:2004th}
S.~W.~Bosch, B.~O.~Lange, M.~Neubert and G.~Paz,
``Factorization and shape-function effects in inclusive B-meson decays,''
[hep-ph/0402094].
 
\bibitem{BDK}
E.~Gardi,
JHEP {\bf 0404}, 049 (2004)
[hep-ph/0403249].


\bibitem{renormalons}
M.~Beneke,
{\em Phys. Rept.}  {\bf 317} (1999) 1;
M.~Beneke and V.~M.~Braun,
``Renormalons and power corrections,'',
in the Boris Ioffe Festschrift, {\it At the Frontier of Particle
Physics / Handbook of QCD}, ed. M. Shifman (World Scientific, Singapore, 2001),
vol. 3, p. 1719 [hep-ph/0010208].


\bibitem{Bigi:1994em}
I.~I.~Y.~Bigi, M.~A.~Shifman, N.~G.~Uraltsev and A.~I.~Vainshtein,
{\em Phys. Rev.} {\bf D50} (1994) 2234
[hep-ph/9402360].

\bibitem{Beneke:1994sw}
M.~Beneke and V.~M.~Braun,
{\em Nucl. Phys.} {\bf B426} (1994) 301
[hep-ph/9402364].

\bibitem{BBZ}
M.~Beneke, V.~M.~Braun and V.~I.~Zakharov,
{\em Phys. Rev. Lett.}  {\bf 73} (1994) 3058
[hep-ph/9405304].

\bibitem{Neubert:1994wq}
M.~Neubert and C.~T.~Sachrajda,
{\em Nucl. Phys.} {\bf B438} (1995) 235
[hep-ph/9407394].


\bibitem{Gardi:2001ny}
E.~Gardi and J.~Rathsman,
{\em Nucl. Phys.}  {\bf B609} (2001) 123
[hep-ph/0103217];
{\em Nucl. Phys.}  {\bf B638} (2002) 243
[hep-ph/0201019].

\bibitem{Gardi:2001di}
E.~Gardi,
{\em Nucl. Phys.} {\bf B622} (2002) 365
[hep-ph/0108222].

\bibitem{Gardi:2002bk}
E.~Gardi, G.~P.~Korchemsky, D.~A.~Ross and S.~Tafat,
{\em Nucl. Phys.} {\bf B636} (2002) 385
[hep-ph/0203161].

\bibitem{Gardi:2002xm}
E.~Gardi and R.~G.~Roberts,
{\em Nucl. Phys.}  {\bf B653} (2003) 227
[hep-ph/0210429].

\bibitem{CG}
M.~Cacciari and E.~Gardi,
{\em Nucl. Phys.} {\bf B664} (2003) 299
[hep-ph/0301047].

\bibitem{Gardi:2003iv}
E.~Gardi and L.~Magnea,
JHEP {\bf 0308} (2003) 030
[hep-ph/0306094].


\bibitem{Hoang:1998nz}
A.~H.~Hoang, M.~C.~Smith, T.~Stelzer and S.~Willenbrock,
{\em Phys.\ Rev.\ }  {\bf D59} (1999) 114014
[hep-ph/9804227].

\bibitem{Berger:2004xf}
C.~F.~Berger and L.~Magnea,
``Scaling of power corrections for angularities from dressed gluon
exponentiation,''
[hep-ph/0407024].

\bibitem{CSS} 
G.~Sterman,
{\em Nucl. Phys.} {\bf B281} (1987) 310;
J.~C.~Collins, D.~E.~Soper and G.~Sterman,
{\em Adv. Ser. Direct. High Energy Phys.}  {\bf 5} (1988) 1, published in
`Perturbative QCD', A.H. Mueller, ed. (World Scientific Publ., 1989). 

\bibitem{CT_CTTW}
S.~Catani and L.~Trentadue,
{\rm Nucl. Phys.}  {\bf B327} (1989) 323;
S. Catani, L. Trentadue, G. Turnock and B.R. Webber,
{\em Nucl. Phys.} {\bf B407} (1993) 3.


\end{thebibliography}
\end{document}